\documentclass[12pt]{article}
\usepackage[dvips]{color}
\usepackage{epsfig}
\usepackage{amsmath}
\usepackage{graphicx}

\begin{document}
\title{Effect of Thermal Fluctuations on a Charged Dilatonic Black Saturn}
\author{{Behnam Pourhassan$^{a}$\thanks{Email: b.pourhassan@du.ac.ir} and Mir Faizal$^{b}$\thanks{Email: f2mir@uwaterloo.ca}}\\
$^{a}${\small {\em School of Physics, Damghan University, Damghan, Iran}}\\
$^{b}${\small {\em  Department of Physics and Astronomy, University of Lethbridge,  Lethbridge, AB T1K 3M4, Canada}}}
\date{}
\maketitle

\begin{abstract}
In this paper, we will analyze the effect of thermal fluctuations on the thermodynamics of a charged dilatonic black Saturn.
These thermal fluctuations will correct the thermodynamics of the charged dilatonic black Saturn. We  will analyze the corrections
to the thermodynamics of this system by first relating the fluctuations in the entropy to the fluctuations in the energy.
Then, we will use the relation between entropy and a conformal field theory to analyze the
fluctuations in the entropy. We will demonstrate that similar physical results are obtained from both these approaches.
We will also study the effect of thermal fluctuations on the phase transition in this charged dilatonic black Saturn.
\end{abstract}

\section{Introduction}
Black Saturn is an interesting black object in higher dimensions, where a black hole is  surrounded by a black ring \cite{0701035, 1007.3668}. The
 black ring and the   black hole are in thermodynamic equilibrium with each other
\cite{0705.3697}. This thermodynamic equilibrium is obtained because of the
rotation of the black ring. However, it is also possible to obtain thermodynamically stable black Saturn solutions with
a  static black ring \cite{0802.0784, 0809.0154}.
In this case, the    black hole   remains in thermodynamic equilibrium with the    static black ring  because of an external magnetic field.
It may be noted that conditions for meta-stability of a black Saturn have also been studied \cite{0804.4575}. It has been demonstrated
that the black Saturn is causal stably on the
closure of the domain of outer communications \cite{1102.3942}. It has been possible to obtain a relation between
the black Saturn and Myers-Perry black holes \cite{1309.4414}.
Furthermore,  the thermodynamics of   charged  black rings and a dilatonic black Saturn has
been analyzed using Einstein-Maxwell-dilaton theory in five dimensions.
\cite{1407.2009}.
In this paper, we will analyze the effects of thermal fluctuations
on the thermodynamics of this charged dilatonic black Saturn.\\

It is possible to study the thermodynamics of black Saturn as an entropy is associated with all black objects. This is needed
to prevent the violation of the second
law of thermodynamics \cite{1, 1a}. This is because the entropy of the universe would spontaneous reduce when an object would crosses the horizon of any
black object,
if we do not associate an entropy with that  black object. So,  in order to prevent the violation of second law of thermodynamics
for all black objects, an entropy is associated with them. In fact,  black holes
have more entropy than any other object with the same volume \cite{2, 4}. This maximum entropy which is associated with a black holes is
proportional to the area of the  horizon \cite{4a}. The fact that this
entropy scales with the area and not the volume of the black holes has lead
to the development of the holographic principle \cite{5, 5a}. According to the holographic principle the degrees of freedom
in a region of space are equal to the degrees of freedom on the boundary surrounding that region of space.
It is expected that this holographic principle will get modified near Planck scale \cite{6, 6a}.\\

It is generally expected that any thermodynamical system will undergo thermal fluctuations. These thermal fluctuations will lead to
corrections in the thermodynamics
of that system. As black holes are also thermodynamic systems, they will also be effected by thermal fluctuations.
In fact, the effect of such thermal fluctuations on the entropy of the black holes has been analyzed \cite{l1, SPR}.
These thermal fluctuations  modify the entropy-area law of the black holes. This is an interesting result as
in Jacobson formalism,
that the Einstein's equation can be derived from the first law of thermodynamics \cite{z12j, jz12}. This is done by requiring that the
Clausius relation holds for all the local Rindler causal horizons through each space-time point. As there exists a relation between the
geometry of black holes and thermodynamics, we expect that thermal fluctuations will also give rise to the fluctuations of the metric
in Jacobson thermodynamic formalism. Hence, we can expect that these thermal fluctuations in the thermodynamic of black holes to occur
because of quantum fluctuations in the geometry of space-time. We can neglect such fluctuations for large black holes.
However, as the black holes reduce in size due to Hawking radiation, the effect of quantum fluctuations on the geometry of the black
holes cannot be neglected.
Thus, at this stage, the effect of thermal fluctuations on the thermodynamics of black holes also cannot be neglected.\\

It is interesting to note that the thermal fluctuations correct the entropy of the black holes by a
logarithmic   terms \cite{l1, SPR}.  This is because such logarithmic corrections   also occur in various
different approaches to quantum gravity.  In fact, non-perturbative quantum  general
relativity has been used for calculating the corrections to the thermodynamics of black holes \cite{1z}. In this approach, the
leading order corrections to the entropy of a black hole are logarithmic corrections. These logarithmic corrections have
been calculated from the density of states of a black hole. This density of states are calculated using the conformal blocks of a well
defined  conformal field theory. Such logarithmic correction   have also been calculated using
the Cardy formula \cite{card}.  The logarithmic correction  for a BTZ have been obtained using an exact
partition function \cite{gks}. Such corrections terms have also been obtained by analyze the effect
of matter fields surrounding a black hole \cite{other, other0, other1}.\\

The string theory considerations have also lead to  logarithmic corrections in the entropy of black holes \cite{solo1, solo2, solo4, solo5}.
In fact, such corrections terms have  been also obtained for dilatonic black holes \cite{jy}. The studies done on the partition
function of a black hole have also lead to such logarithmic correction   \cite{bss}.
The generalized uncertainty principle has also been used for calculating the corrections to the thermodynamics of a black hole
\cite{mi, r1}. The entropy of a black hole gets modified by the generalized uncertainty principle. The
correction terms obtained this way can be expressed as logarithmic functions of the area of the horizon.\\

It may be noted that quantum fluctuations become important for all black geometries, when the size of the such geometries is sufficiently reduced.
So, at such small scales the thermal fluctuations also become important for  all black geometries.
The effect of thermal fluctuations on the thermodynamics of an AdS charged black hole has already been analyzed \cite{1503.07418}.
This was done by relating the fluctuations in the entropy of this black hole to a conformal field theory.
Furthermore, the effect of thermal fluctuations on the thermodynamics of a black Saturn has also been studied \cite{1505.02373}.
This was done by relating the fluctuations in the entropy of the black Saturn to the   fluctuations in its energy.
However, so far no work has been done on the effect of thermal fluctuations on the thermodynamics of a charged  dilatonic black Saturn.
So, in this paper, we will analyze the effects of thermal fluctuations  on the thermodynamics of a charged  dilatonic black Saturn.
We will first  analyze the  thermal fluctuations of the charged  dilatonic black Saturn by relating the fluctuations in the entropy
of the charged  dilatonic black Saturn
to the fluctuations in its energy. Then we will  use the relation between the fluctuations in the entropy and a conformal field theory
to analyze the effect of such thermal fluctuations. We will demonstrate that both these approaches lead to the same
physical effects for the  charged
dilatonic black Saturn.

\section{Charged Dilatonic Black Saturn}
In this section, we review some basic properties of charged dilatonic black Saturn   \cite{1407.2009}.
The metric for black Saturn can be written as \cite{1407.2009}
\begin{eqnarray}\label{B1}
ds^{2}&=&-V_{\beta}(\rho,z)^{-\frac{2}{3}}\frac{H_{y}}{H_{x}}\left[dt+(\frac{\omega_{\psi}}{H_{y}}+q)d\psi\right]^{2}\nonumber\\
&+&V_{\beta}(\rho,z)^{-\frac{1}{3}}H_{x}\left[k^{2}P(d\rho^{2}+dz^{2})+\frac{G_{y}}{H_{y}}d\psi^{2}+\frac{G_{x}}{H_{x}}d\varphi^{2}\right],
\end{eqnarray}
where $q$ and $k$ are constants, and $\beta$ is related to the charge of the black Saturn. Furthermore, we also have
\begin{equation}\label{B1-2}
V_{\beta}(\rho,z)=\cosh(\beta)^{2}-\frac{H_{y}}{H_{x}}\sinh(\beta)^{2},
\end{equation}
and
\begin{eqnarray}\label{B2}
G_{x}&=&\frac{\mu_{4}}{\mu_{3}\mu_{5}}\rho^{2}\nonumber\\
G_{y}&=&\frac{\mu_{3}\mu_{5}}{\mu_{4}}.
\end{eqnarray}
Here, we have used
\begin{equation}\label{B3}
P=(\mu_{3}\mu_{4}+\rho^{2})^{2}(\mu_{1}\mu_{5}+\rho^{2})(\mu_{4}\mu_{5}+\rho^{2}),
\end{equation}
and
\begin{equation}\label{B4}
\mu_{i}=\sqrt{\rho^{2}+(z-a_{i})^{2}}-(z-a_{i})=R_{i}-(z-a_{i}).
\end{equation}
Here  $a_{i}$ ($i=1,...,5$)  are real constant parameters which satisfy
\begin{equation}\label{B5}
a_{1}\leq a_{5}\leq a_{4}\leq a_{3}\leq a_{2}.
\end{equation}
The non-zero components of the vector potential are given by,
\begin{eqnarray}\label{B5-2}
A_{t}&=&\frac{(H_{x}-H_{y})\sinh(\beta)\cosh(\beta)}{H_{x}\cosh(\beta)^{2}-H_{y}\sinh(\beta)^{2}},\nonumber\\
A_{\psi}&=&\frac{(\omega_{\psi}+qH_{y})\sinh(\beta)}{H_{y}\sinh(\beta)^{2}-H_{x}\cosh(\beta)^{2}}
\end{eqnarray}
and  the dilaton function is given by,
\begin{equation}\label{B5-3}
\Phi=-\frac{\sqrt{6}}{3}\ln\left(\cosh(\beta)^{2}-\frac{H_{y}}{H_{x}}\sinh(\beta)^{2}\right).
\end{equation}
Furthermore, we also have
\begin{eqnarray}\label{B6}
H_{x}&=&\frac{M_{0}+c_{1}^{2}M_{1}+c_{2}^{2}M_{2}+c_{1}c_{2}M_{3}+c_{1}^{2}c_{2}^{2}M_{4}}{F}\nonumber\\
H_{y}&=&\frac{1}{F}\frac{\mu_{3}}{\mu_{4}}
\left[\frac{\mu_{1}}{\mu_{2}}M_{0}-c_{1}^{2}M_{1}\frac{\rho^{2}}{\mu_{1}\mu_{2}}-c_{2}^{2}M_{2}
\frac{\mu_{1}\mu_{2}}{\rho^{2}}
+c_{1}c_{2}M_{3}+c_{1}^{2}c_{2}^{2}M_{4}\frac{\mu_{2}}{\mu_{1}}\right],
\end{eqnarray}
where $c_{1}$ and $c_{2}$ are real constants, and
\begin{eqnarray}\label{B7}
M_{0}&=&\mu_{2}\mu_{5}^{2}(\mu_{1}-\mu_{3})^{2}(\mu_{2}-\mu_{4})^{2}(\rho^{2}+
\mu_{1}\mu_{2})^{2}(\rho^{2}+\mu_{1}\mu_{4})^{2}(\rho^{2}+\mu_{2}\mu_{3})^{2},\nonumber\\
M_{1}&=&\mu_{1}^{2}\mu_{2}\mu_{3}\mu_{4}\mu_{5}\rho^{2}(\mu_{1}-\mu_{2})^{2}
(\mu_{2}-\mu_{4})^{2}(\mu_{1}-\mu_{5})^{2}(\rho^{2}+\mu_{2}\mu_{3})^{2},\nonumber\\
M_{2}&=&\mu_{2}\mu_{3}\mu_{4}\mu_{5}\rho^{2}(\mu_{1}-\mu_{2})^{2}
 (\mu_{1}-\mu_{3})^{2}(\rho^{2}+\mu_{1}\mu_{4})^{2}(\rho^{2}+\mu_{2}\mu_{5})^{2},\nonumber\\
M_{3}&=&2\mu_{1}\mu_{2}\mu_{3}\mu_{4}\mu_{5}(\mu_{1}-\mu_{3})(\mu_{1}-\mu_{5})
(\mu_{2}-\mu_{4})(\rho^{2}+\mu_{1}^{2})(\rho^{2}+\mu_{2}^{2})\nonumber \\ && \times
(\rho^{2}+\mu_{1}\mu_{4})(\rho^{2}+\mu_{2}\mu_{3})(\rho^{2}+\mu_{2}\mu_{5}),\nonumber\\
M_{4}&=&\mu_{1}^{2}\mu_{2}\mu_{3}^{2}\mu_{4}^{2}(\mu_{1}-\mu_{5})^{2}(\rho^{2}+\mu_{1}
\mu_{2})^{2}(\rho^{2}+\mu_{2}\mu_{5})^{2},\nonumber\\
\end{eqnarray}
with
\begin{eqnarray}\label{B8}
F&=&\mu_{1}\mu_{5}(\mu_{1}-\mu_{3})^{2}(\mu_{2}-\mu_{4})^{2}(\rho^{2}+\mu_{1}\mu_{3})\nonumber \\ && \times
(\rho^{2}+\mu_{2}\mu_{3})(\rho^{2}+\mu_{1}\mu_{4})(\rho^{2}+\mu_{2}\mu_{4})
(\rho^{2}+\mu_{2}\mu_{5})\nonumber\\
&&\times(\rho^{2}+\mu_{3}\mu_{5})(\rho^{2}+\mu_{1}^{2})(\rho^{2}+\mu_{2}^{2})(\rho^{2}+
\mu_{3}^{2})(\rho^{2}+\mu_{4}^{2})(\rho^{2}+\mu_{5}^{2}).
\end{eqnarray}
Here $\omega_{\psi}$ is expressed as
\begin{equation}\label{B9}
\omega_{\psi}=\frac{2}{F\sqrt{G_{x}}}\left[c_{1}R_{1}\sqrt{M_{0}M_{1}}-c_{2}R_{2}
\sqrt{M_{0}M_{2}}+c_{1}^{2}c_{2}R_{2}\sqrt{M_{1}M_{4}}-c_{1}c_{2}^{2}R_{1}\sqrt{M_{2}M_{4}}\right],
\end{equation}
where $R_{1}$ and $R_{2}$ given by the relation (\ref{B4}).
Free parameters of this model are fixed as \cite{1007.3668},
\begin{equation}\label{B10}
L^{2}=a_{2}-a_{1},
\end{equation}
and
\begin{equation}\label{B11}
c_{1}=\pm\sqrt{\frac{2(a_{3}-a_{1})(a_{4}-a_{1})}{a_{5}-a_{1}}}.
\end{equation}
We also have
\begin{equation}\label{B12}
c_{2}=\sqrt{2}(a_{4}-a_{2})\frac{\sqrt{(a_{1}-a_{3})(a_{4}-a_{2})(a_{2}-a_{5})(a_{3}-a_{5})}
\pm(a_{2}-a_{1})(a_{3}-a_{4})}
{\sqrt{(a_{1}-a_{4})(a_{2}-a_{4})(a_{1}-a_{5})(a_{2}-a_{5})(a_{3}-a_{5})}},
\end{equation}
and
\begin{equation}\label{B13}
k=\frac{2(a_{1}-a_{3})(a_{2}-a_{4})}{2(a_{1}-a_{3})(a_{2}-a_{4})+(a_{1}-a_{5})c_{1}c_{2}}=
\frac{2k_{1}\hat{k}_{2}}{2k_{1}\hat{k}_{2}+c_{1}c_{2}k_{3}},
\end{equation}
where,
\begin{equation}\label{B14}
\hat{k}_{i}=1-k_{i}=1-\frac{a_{i+2}-a_{1}}{L^{2}},
\end{equation}
with $i=1,2,3$. Here we have
\begin{equation}\label{B5}
0\leq k_{3}\leq k_{2}\leq k_{1}\leq 1.
\end{equation}
The variable $q$ can be written as
\begin{equation}\label{B15}
q=\frac{2k_{1}c_{2}}{2k_{1}-2k_{1}k_{2}+c_{1}c_{2}k_{3}}.
\end{equation}
Thus, all the  thermodynamics quantities can be written in terms of $a_{i}$ with $i=1,2,3,4,5$.\\
The Hawking temperatures for the charged dilatonic black Saturn is given by \cite{1007.3668},
\begin{eqnarray}\label{T01}
T&=&\frac{1}{2\pi L\cosh(\beta)}\sqrt{\frac{\hat{k}_{2}\hat{k}_{3}}{2\hat{k}_{1}}}\left(\frac{(1+k_{2}c)^{2}}
{1+\frac{k_{1}k_{2}\hat{k}_{2}\hat{k}_{3}}{k_{3}\hat{k}_{1}}c^{2}}\right)\nonumber\\
&+&\frac{1}{2\pi L\cosh(\beta)}\sqrt{\frac{k_{1}\hat{k}_{3}(k_{1}-k_{3})}{2k_{2}(k_{2}-k_{3})}}
\left(\frac{(1+k_{2}c)^{2}}{1-(k_{1}-k_{2})c+\frac{k_{1}k_{2}\hat{k}_{3}}{k_{3}}c^{2}}\right),
\end{eqnarray}
where
\begin{equation}\label{T02}
c=\frac{1}{k_{2}}\left(\varepsilon\frac{k_{1}-k_{2}}{\sqrt{k_{1}\hat{k}_{2}\hat{k}_{3}(k_{1}-k_{3})}}-1\right).
\end{equation}
Here we have used $\varepsilon=\pm1$. It may be noted that $\varepsilon=0$  gives a naked singularity.
The entropy of the charged dilatonic black Saturn, in absence of thermal fluctuations,  is given by
\begin{eqnarray}\label{T03}
S_{0}&=&\frac{\pi^{2} L^{3}\cosh(\beta)}{(1+k_{2}c)^{2}}\sqrt{\frac{2\hat{k}_{1}^{3}}{\hat{k}_{2}\hat{k}_{3}}}
\left(1+\frac{k_{1}k_{2}\hat{k}_{2}\hat{k}_{3}c^{2}}{k_{3}\hat{k}_{1}}\right)\nonumber\\
&+&\frac{\pi^{2} L^{3}\cosh(\beta)}{(1+k_{2}c)^{2}}\sqrt{\frac{2k_{2}(k_{2}-k_{3})^{3}}{k_{1}(k_{1}-k_{3})
\hat{k}_{3}}}\left(1-(k_{1}-k_{2})c+\frac{k_{1}k_{2}\hat{k}_{3}c^{2}}{k_{3}}\right).
\end{eqnarray}
The ADM mass of the charged dilatonic black Saturn is given by,
\begin{equation}\label{T04}
 M_{ADM}=\frac{3\pi L^{2}\left(1+\frac{2}{3}\sinh(\beta)^{2}\right)\left(k_{3}
(\hat{k}_{1}+k_{2})-2k_{2}k_{3}(k_{1}-k_{2})c+k_{2}[k_{1}-k_{2}k_{3}(\hat{k}_{2}+k_{1})]c^{2}\right)}{4k_{3}(1+k_{2}c)^{2}},
\end{equation}
which can interpreted as enthalpy $H=M_{ADM}$ \cite{4a,Dolan1,JJP}.
The results for the ordinary black Saturn are recovered when  $\beta=0$ \cite{1505.02373}.
Now if   $a_{i}$ are depend to each other, then  one can express  $a_{i}$ in terms of $a_{1}$. So,    all the
thermodynamical relation can express in terms of $a_{1}$.   It is possible to write
\begin{eqnarray}\label{T04-1}
a_{2}&=&120a_{1}^{2},\nonumber\\
a_{3}&=&24a_{1}^{2},\nonumber\\
a_{4}&=&6a_{1}^{2},\nonumber\\
a_{5}&=&2a_{1}^{2},
\end{eqnarray}
such that the   condition (\ref{B5}) is satisfied.
Even though this  is a special choice,  and we can show that, for any other choice of $a_{i}=\alpha_{i}a_{1}^{2x}$
($i=2,3,4,5$ and $x=1,2,3,\cdots$), we will get similar results.  It is useful to define the
following functions,
\begin{equation}\label{T05}
f(a_{1})=\frac{1}{L}\sqrt{\frac{\hat{k}_{2}\hat{k}_{3}}{2\hat{k}_{1}}}\left(\frac{(1+k_{2}c)^{2}}
{1+\frac{k_{1}k_{2}\hat{k}_{2}\hat{k}_{3}}{k_{3}\hat{k}_{1}}c^{2}}\right),
\end{equation}
and
\begin{equation}\label{T06}
g(a_{1})=\frac{1}{L}\sqrt{\frac{k_{1}\hat{k}_{3}(k_{1}-k_{3})}{2k_{2}(k_{2}-k_{3})}}
\frac{(1+k_{2}c)^{2}}{1-(k_{1}-k_{2})c+\frac{k_{1}k_{2}\hat{k}_{3}}{k_{3}}c^{2}}.
\end{equation}
In the Fig. 1, we give plots of $f$, $g$ and $f+g$ in terms of $a_{1}$.
Surprisingly, we can see that $f+g$ behaves like $g$, so we can write  $f+g\approx g$.

\begin{figure}[h!]
 \begin{center}$
 \begin{array}{cccc}
\includegraphics[width=50 mm]{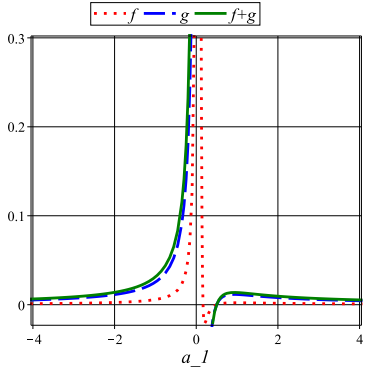}
 \end{array}$
 \end{center}
\caption{$f$, $g$, and $f+g$ in terms of $a_{1}$.}
 \label{fig:1}
\end{figure}

\begin{figure}[h!]
 \begin{center}$
 \begin{array}{cccc}
\includegraphics[width=50 mm]{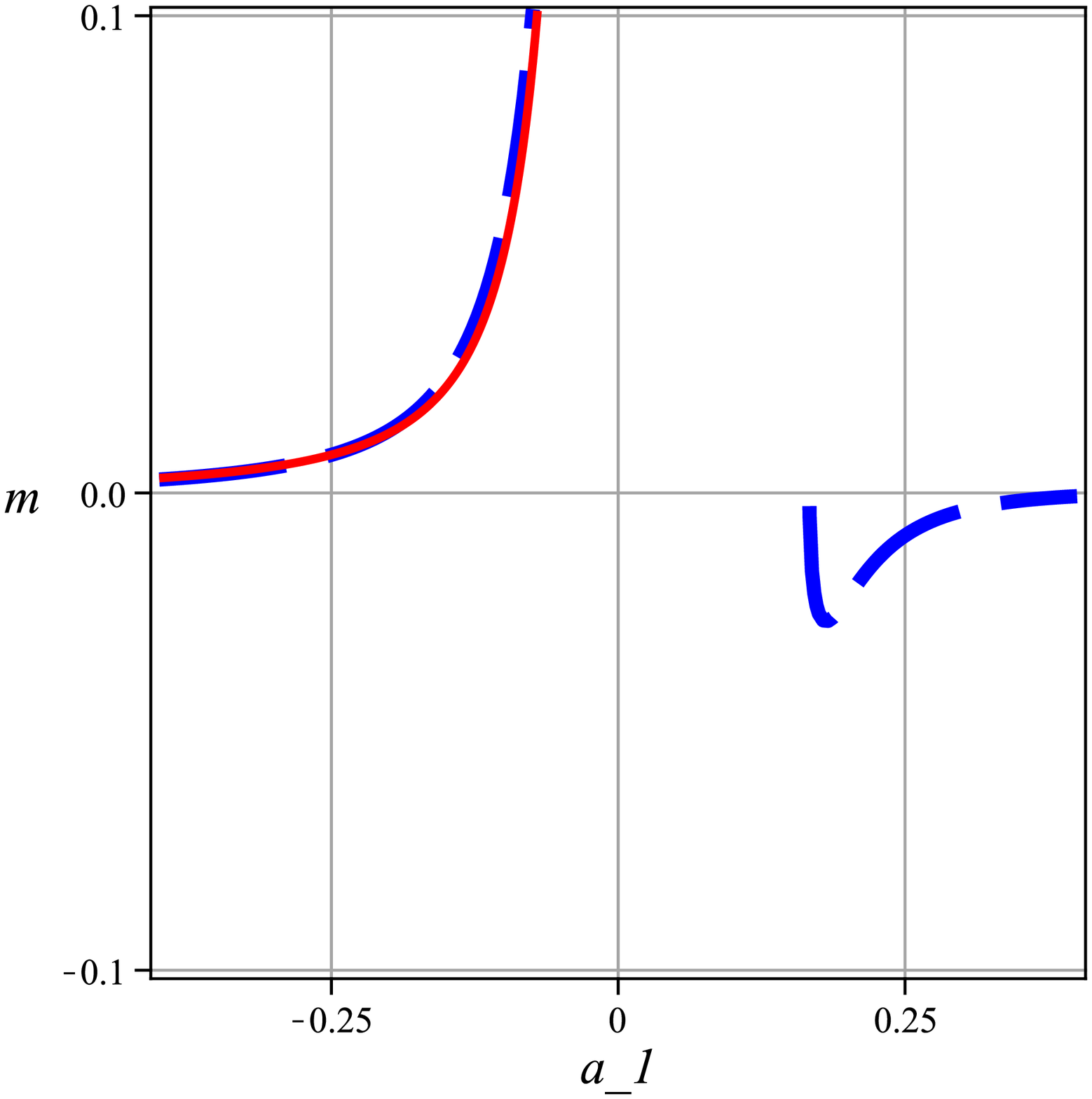}&\includegraphics[width=50 mm]{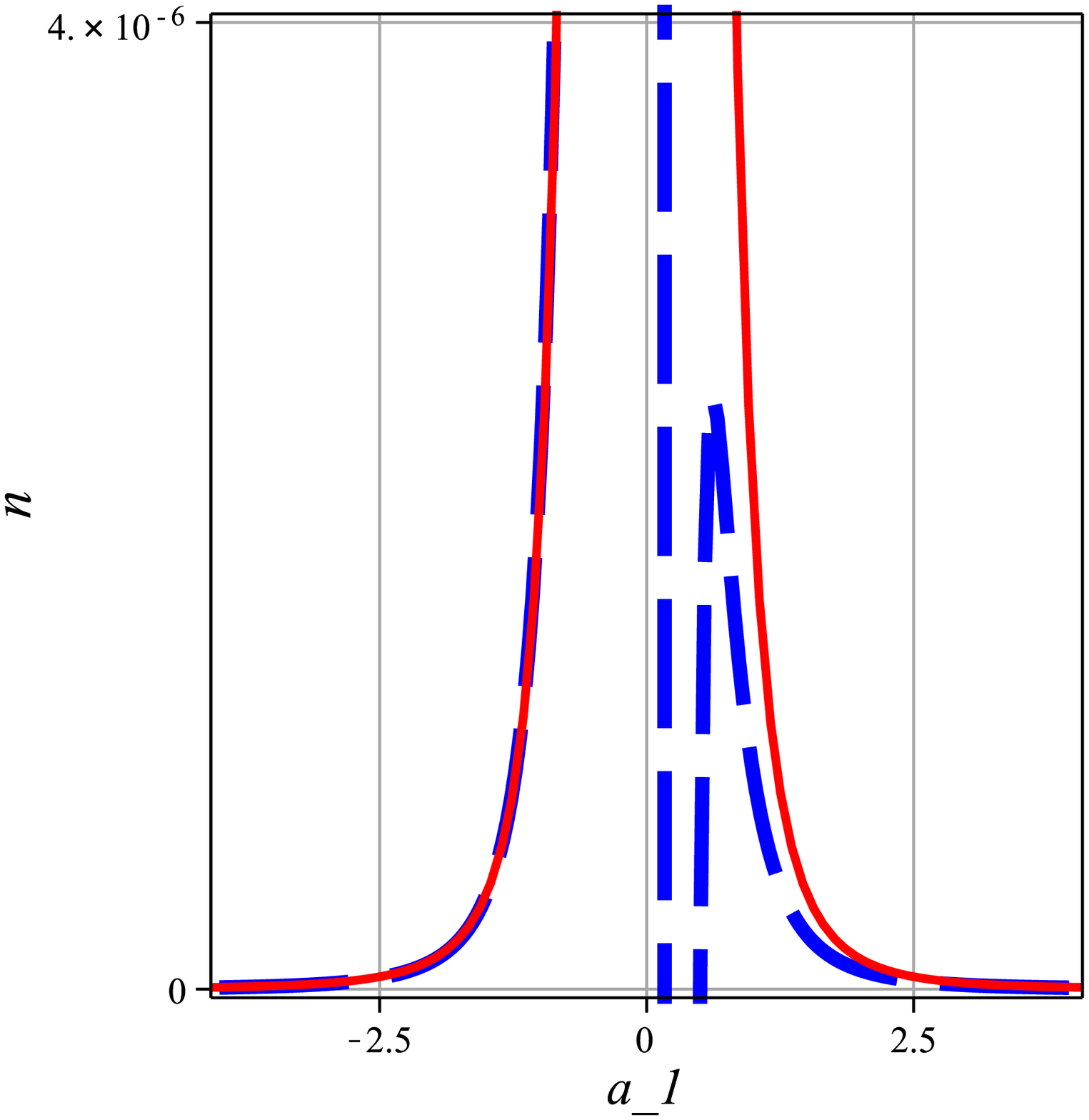}
 \end{array}$
 \end{center}
\caption{$m$ and $n$ in terms of $a_{1}$. Blue dashed lines represent $m$ and $n$ given by the equation (\ref{T09}) while red solid lines
represent fitted functions.}
 \label{fig:2}
\end{figure}

On the other hand, from the fact that $\delta\equiv\frac{4a_{1}^{2}}{a_{1}(24a_{1}-1)}\ll1$
(for small $a_{1}$), we have $\frac{1}{f}+\frac{\delta}{g}\approx\frac{1}{f}$. So, we can write
\begin{equation}\label{T07}
T=\frac{m(a_{1})}{\cosh{\beta}},
\end{equation}
and
\begin{equation}\label{T08}
S_{0}=\frac{\cosh{\beta}}{n(a_{1})},
\end{equation}
with,
\begin{eqnarray}\label{T09}
m(a_{1})&=&\frac{g}{2\pi\sqrt{a_{1}(120a_{1}-1)}},\nonumber\\
n(a_{1})&=&\frac{f}{\pi^{2} a_{1}(24a_{1}-1)\sqrt{a_{1}(120a_{1}-1)}},
\end{eqnarray}
where $f$ and $g$ are given by   (\ref{T05}) and (\ref{T06}). So, we obtain  simple expressions
for the temperature and the entropy of this black Saturn. In the Fig. \ref{fig:2}, we give plots of $m$ and $n$. We find that for
  the temperature
to be positive,  $T\geq0$,    $a_{1}$ should be negative. This is because from the  left plot of the Fig.
\ref{fig:2}, if $a_{1}>0$, then $m<0$, and so   $T<0$. As the temperature cannot be negative,
so we obtain the first   constraint on $a_1$ i.e.,  $a_{1}<0$.
Thus, the physical results are restricted to the    left side of each plots ($a_{1}<0$). In this region,
 $m$ and $n$ can be expressed using simple functions.
As it illustrated by solid red lines of the Fig. \ref{fig:2}, the function $m$ with negative $a_{1}$ behaves as $\frac{0.0005}{a_{1}^{2}}$,
while the function $n$ with negative $a_{1}$ behaves as $\frac{0.000002}{a_{1}^{4}}$. Coefficients $0.0005$ and $0.000002$   are obtained using the
specific  choice of coefficients chosen in (\ref{T04-1}). If we change power (to any even power) or even change coefficients defined in (\ref{T04-1}),
then we can fix function by changing the value of $0.0005$ and $0.000002$.\\

\begin{figure}[h!]
 \begin{center}$
 \begin{array}{cccc}
\includegraphics[width=50 mm]{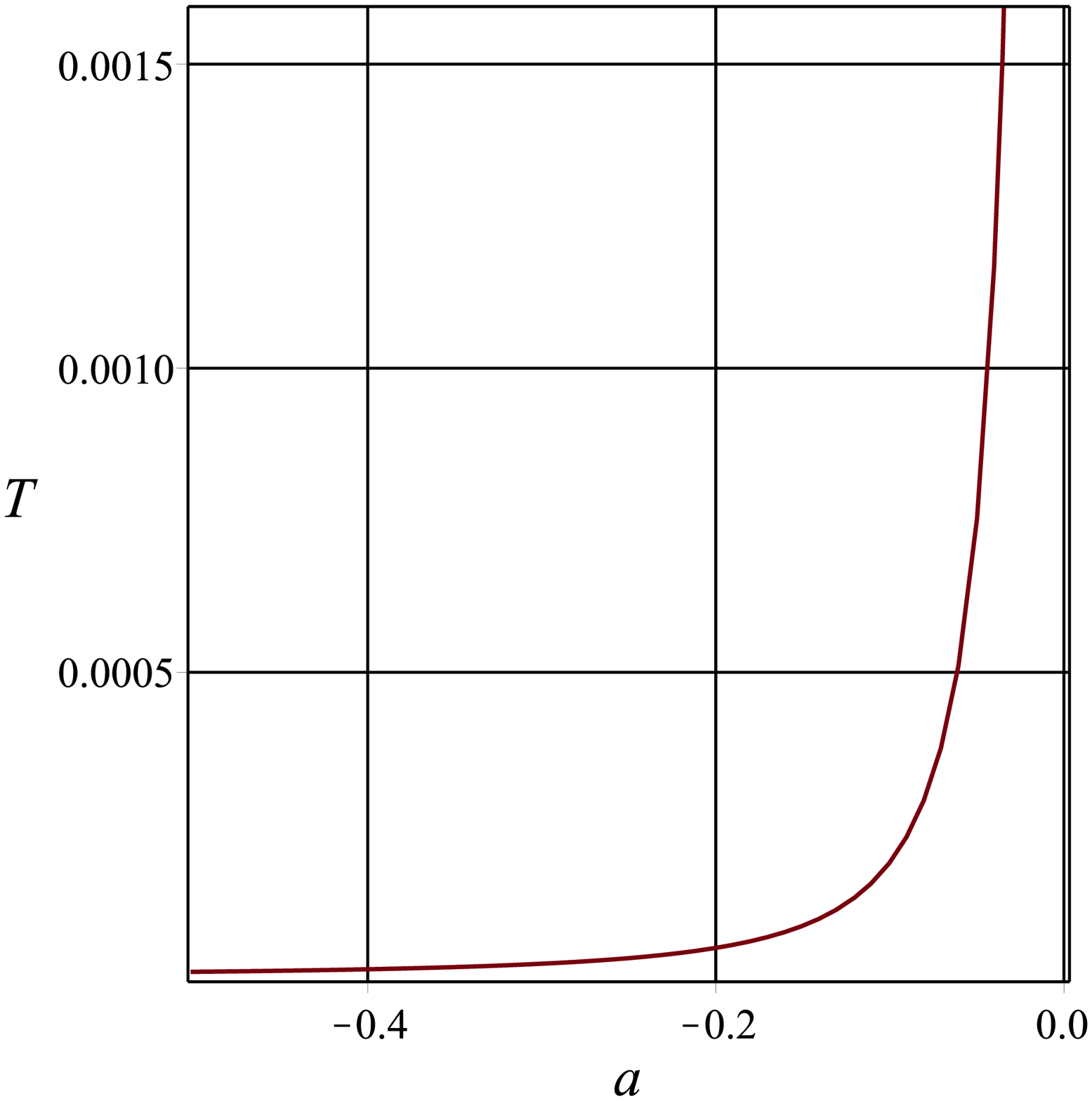}&\includegraphics[width=50 mm]{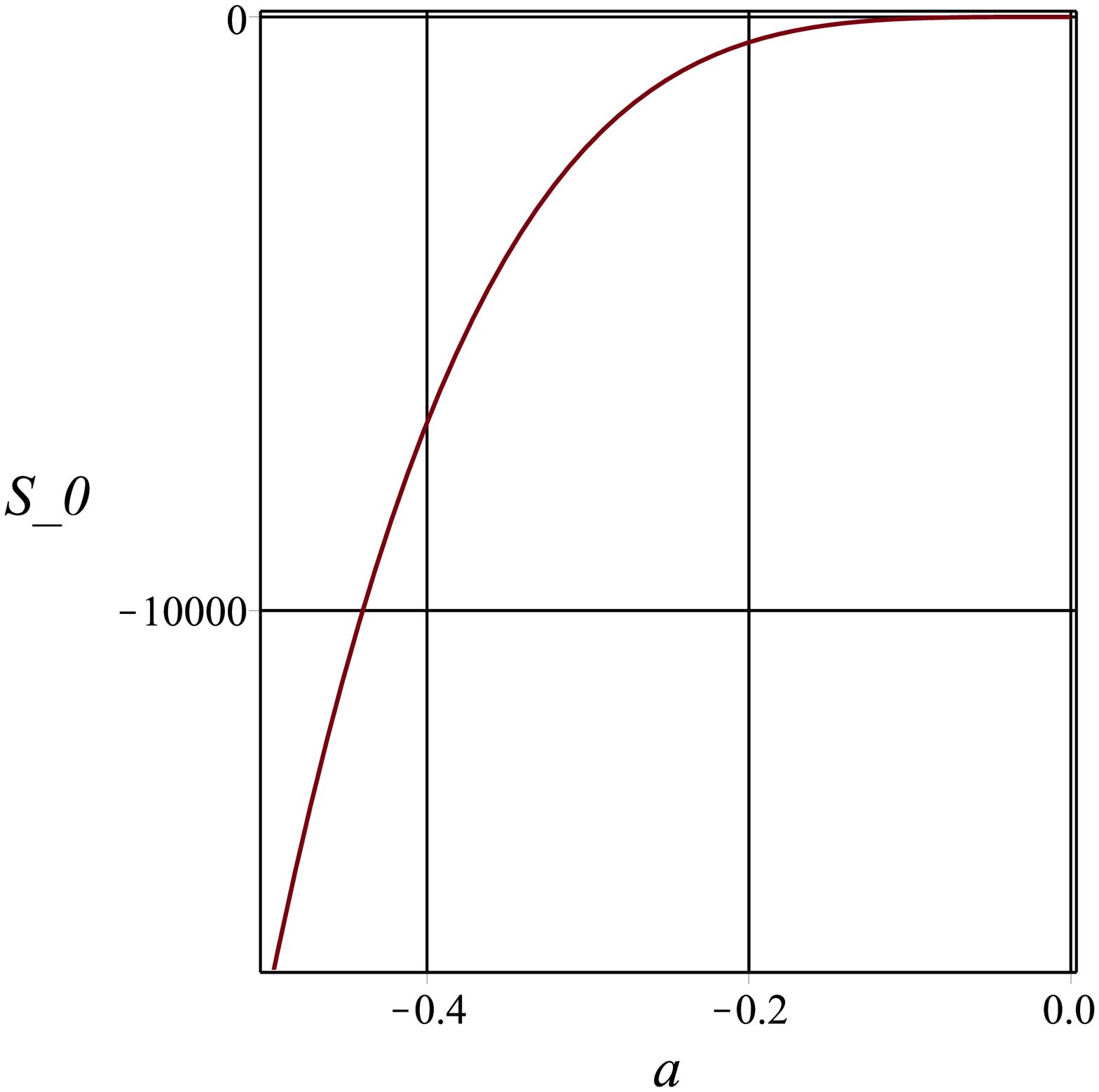}
 \end{array}$
 \end{center}
\caption{Typical behavior of $T$ and $S_{0}$ in terms of $a_{1}$ for arbitrary $\beta$, $\delta_{1}=0.0005$ and $\delta_{2}=0.000002$.}
 \label{fig:new1}
\end{figure}

Hence, we have a very simple expression for the temperature and  entropy  of this black Saturn,
\begin{equation}\label{T10}
T=\frac{\delta_{1}}{a_{1}^{2}\cosh{\beta}},
\end{equation}
and
\begin{equation}\label{T11}
S_{0}=\frac{a_{1}^{4}\cosh{\beta}}{\delta_{2}},
\end{equation}
Here  $\delta_{1}=0.0005$ and $\delta_{2}=0.000002$, and these values are obtained using the choice of
coefficient defined in  (\ref{T04-1}). However, these solutions are
general, and they hold for  any choice of $a_{i}=\alpha_{i}a_{1}^{2x}$. The only thing which will change for
a different choice of $a_i$, is the value of $\delta_{1}$ and $\delta_{2}$. Hence,  we can  consider general solution with arbitrary
$\delta_{1}$ and $\delta_{2}$. It is clear that  for the temperature to be positive, $\delta_{1}$ has to be positive.
In the Fig. \ref{fig:new1},  we can see general behavior of $T$ and $S_{0}$. It is illustrated that,
temperature and entropy are increasing function of $a_{1}$.

\section{Energy Fluctuations}
\begin{figure}[h!]
 \begin{center}$
 \begin{array}{cccc}
\includegraphics[width=50 mm]{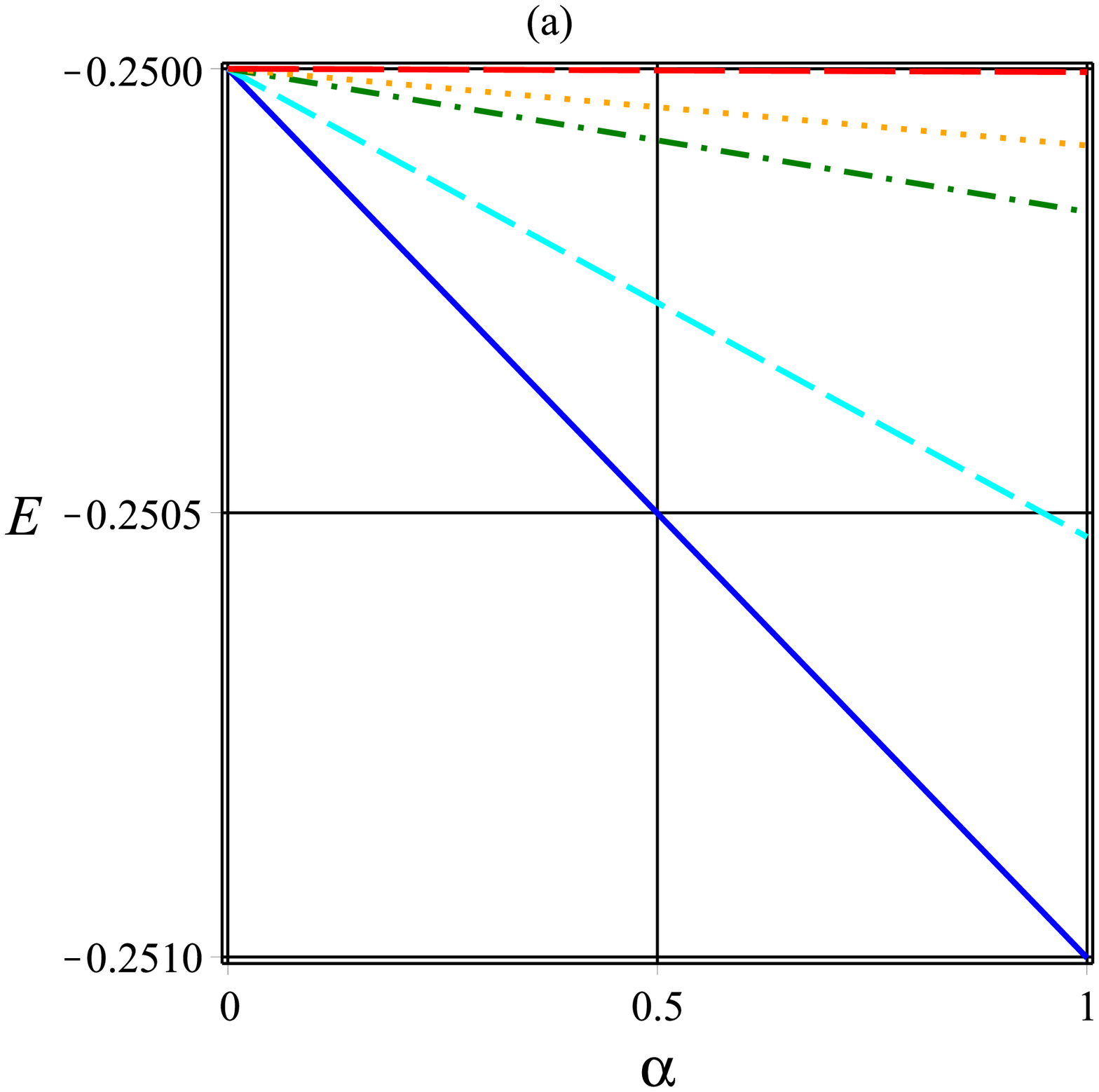}&\includegraphics[width=50 mm]{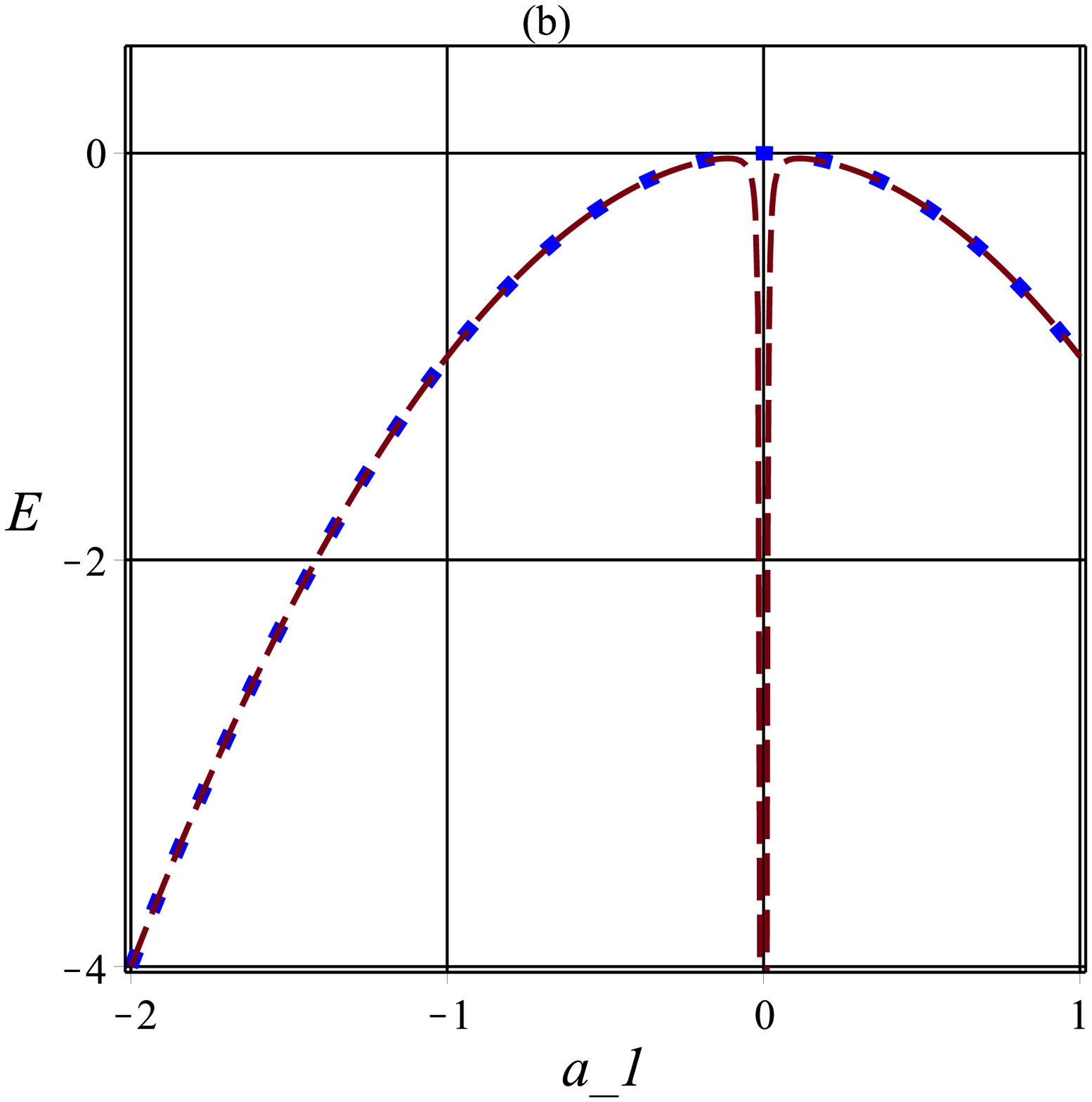}
 \end{array}$
 \end{center}
\caption{$E$ in terms of $\alpha$ (a) and $a_{1}$ (b), with $\delta_{1}=0.0005$, $\delta_{2}=-0.001$. (a) $a_{1}=-0.5$,
$\beta=0$ (blue solid), $\beta=\frac{\pi}{10}$ (cyan dashed), $\beta=\frac{\pi}{5}$ (green dash dotted), $\beta=\frac{\pi}{2}$
(orange dotted), $\beta=2\pi$ (red long dashed). (b) $\beta=1$, $\alpha=0$ (blue dotted), $\alpha=1$ (red solid).}
 \label{fig:3}
\end{figure}

The thermal fluctuations correct the thermodynamics of all black objects. This happens because they correct the partition function for these
black objects. Thus, various different thermodynamical quantities get corrected because of these thermal fluctuations. It is interesting to note that
the entropy of these black objects get corrected by a logarithmic term. So, if $\beta_{\kappa}^{-1} = T$ is the   temperature for the system close to
equilibrium, and $\beta_0^{-1} = T_0 $ is the equilibrium temperature of the system, then we have
$
 S = S_0 -    (\ln  S_0'')/2 $ where $ S_0'' =  ({\partial^2 S}/{\partial \beta_{\kappa}^2} )|_{\beta_{\kappa} = \beta_0}
$.
By using the fact that this second derivative of the entropy can be expressed   in terms of the fluctuation  of energy near the equilibrium,
this expression for the total entropy can be written as  \cite{SPR, 1505.02373},
\begin{equation}\label{TF1}
S = S_{0} -\frac{\alpha}{2} \ln |C_{0}T^{2}|+\cdots,
\end{equation}
where $\alpha=0$ or $\alpha=1$. Furthermore,  almost all different approaches to quantum gravity generate the logarithmic correction term,
but the coefficient of this term varies between different  approaches, it is useful to keep this analysis general and introduce a general parameter
$\alpha$. Here $\alpha = 1$ indicates that we have taken the thermal fluctuations into account, and this hold for a very small black object.
The value $\alpha =0$ indicates that we have not taken thermal fluctuations into account, and this holds for large black objects.
Here   $S_0$ is the original entropy of the charged dilatonic black Saturn given by the equation (\ref{T03}),
and a variable $\alpha$ is introduced to parameterize the effect of thermal fluctuations on the thermodynamics
of charged dilatonic black Saturn. Furthermore, we have also defined
\begin{equation}\label{TF3}
C_{0}=T\frac{\partial S_{0}}{\partial T}.
\end{equation}
  Using the equations (\ref{T10}), (\ref{T11}) and (\ref{TF3}),  we can obtain,
\begin{equation}\label{T11-1}
C_{0}=-2\frac{a_{1}^{4}\cosh{\beta}}{\delta_{2}},
\end{equation}
which means that $\delta_{2}$ should be negative to have thermodynamical stability.
It is possible if we choose some negative coefficients in the relation (\ref{T04-1}).
For example, by choosing $a_{2}=120a_{1}^{2}$, $a_{3}=-24a_{1}^{2}$, $a_{4}=6a_{1}^{2}$, $a_{5}=-2a_{1}^{2}$,
one can obtain $\delta_{2}\approx-0.001$. However, the
positive and the negative  values  $\delta_{2}$ do not effect the form  of the   logarithmic correction (\ref{TF1}).\\
Now using equation (\ref{TF1}),  one can write the internal energy as
\begin{equation}\label{TF4}
E=\int{T dS}=E_{0}-\frac{\alpha}{2}T-\frac{\alpha}{4}T^{2},
\end{equation}
where
\begin{equation}\label{TF5}
E_{0}=\int{T dS_{0}}.
\end{equation}
This internal energy can be expressed in terms of  $a_{1}$,
\begin{equation}\label{TF5-1}
E=\frac {\delta \left( 8\,{a_{1}}^{6}\cosh^{2}{\beta}-2\alpha\delta_{2}\,{a_{1}}^{2}\cosh{\beta}
-\alpha\delta_{1}\delta_{2} \right) }{4\delta_{2}{a_{1}}^{4}
\cosh^{2}{\beta}}.
\end{equation}
We find that for any values of parameter,  energy is decreasing function
of $\alpha$ when $\delta_{1}$ is positive, and it is an increasing function of $\alpha$ when    $\delta_{1}$ is negative. However,
as $\delta_1$ cannot be negative because
  $T\geq0$, so the inner energy cannot be an increasing function of $\alpha$. In the Fig. \ref{fig:3},
we can observe the variation of $E$ for various values $\beta$. For the   values of $\beta$,  such as $10\pi$,  the value of energy is
$E = -0.25$. However, for the larger values of $\beta$, the energy becomes a constant.

It is possible to check validity of the first law of thermodynamics,
\begin{equation}\label{TF6}
dE=TdS+ A da_{1} + B d{\beta},
\end{equation}
where $A$ and $B$ are the thermodynamic variables dual to $a_{1}$ and $\beta$, respectively. 
We will check validity of this equation with the geometro-thermodynamics method at end of this paper.
In the case of $A=0$ and $B=0$, we have ordinary object, where $T=\frac{dE}{dS}$. The corrected entropy can be written as,
\begin{equation}\label{TF7}
S={\frac {{a_{1}}^{4}\cosh{\beta} }{\delta_{2}}}-\frac{1}{2}\alpha \ln
 \left( {\frac {{\delta_{1}}^{2}}{\delta_{2}\cosh{\beta} }}
 \right).
\end{equation}
Also, the Helmholtz free energy is given by
\begin{eqnarray}\label{TF8}
F=E-TS=&-&{\frac {\delta_{1}}{{a_{1}}^{2}\cosh{\beta}} \left( {\frac {{a_{1}
}^{4}\cosh{\beta} }{\delta_{2}}}-\frac{\alpha}{2}\ln  \left( -2{\frac {{\delta_{1}}^{2}}{\delta_{2}\cosh{\beta} }} \right)
 \right) }\nonumber\\
 &+&2\frac{\delta_{1}}{\delta_{2}}a_{1}^{2}-{\frac {\alpha
\delta_{1}}{2{a_{1}}^{2}\cosh{\beta} }}-{\frac {\alpha{\delta_{1}}^{2}}{4{a_{1}}^{4} \cosh^{2}{\beta}}}.
\end{eqnarray}
After some calculations, we obtain
\begin{equation}\label{TF9}
F=F_{0}-\frac{\alpha}{2}T(1-\ln{C_{0}T^{2}})-\frac{\alpha}{4}T^{2},
\end{equation}
where
\begin{equation}\label{TF10}
F_{0}=E_{0}-TS_{0}.
\end{equation}
Finally, the specific heat at constant volume,
\begin{equation}\label{TF11}
C=T\frac{\partial S}{\partial T},
\end{equation}
can be expressed as
\begin{equation}\label{TF12}
C=C_{0}-\frac{\alpha}{2}(1+T),
\end{equation}
where $C_{0}$ given by the equation (\ref{TF3}). Using the equation (\ref{TF11}), it is easy to find that,
\begin{equation}\label{TF12-1}
C= C_{0}=-\frac{2a_{1}^{4}\cosh{\beta}}{\delta_{2}}.
\end{equation}
So, the logarithmic  corrections do not have  any effect on the specific heat. Furthermore,  with negative $\delta_{2}$,
the specific heat is always positive, and so we have thermodynamical stability. In fact, our approximations lead to
$S_{0}T^{2}=C_{0}T^{2}=constant$. In order to analyse the effect of logarithmic corrections, we can use equations (\ref{T10})
and (\ref{T11-1}),  to find the specific heat in terms of temperature,
\begin{equation}\label{TF12-2}
C=-\frac{2\delta_{1}a_{1}^{2}}{T\delta_{2}}-\frac{\alpha}{2}(1+T).
\end{equation}
For the negative $\delta_{2}$, the first term is positive while the second term is negative, so thermodynamical stability needs,
\begin{equation}\label{TF12-3}
T^{2}+T-t\leq0,
\end{equation}
where $c=|4\frac{\delta_{1}a_{1}^{2}}{\delta_{2}\alpha}|$ is  a positive quantity.  So, we can find the critical temperature,
\begin{equation}\label{TF12-4}
T_{c}=\frac{1}{2}[\sqrt{1+4t}-1],
\end{equation}
Now for $T\leq T_{c}$,  we have thermodynamical stability. If $t\leq1$, then  $T_{c}\simeq t$.\\
It is important to note that when    $C_{0}<0$ in $(\ref{TF1})$, the   unperturbed black Saturn is
unstable, as $C_0$ is the specific heat of the unperturbed solution.
\subsection{Conformal Field Theory}
\begin{figure}[h!]
 \begin{center}$
 \begin{array}{cccc}
\includegraphics[width=50 mm]{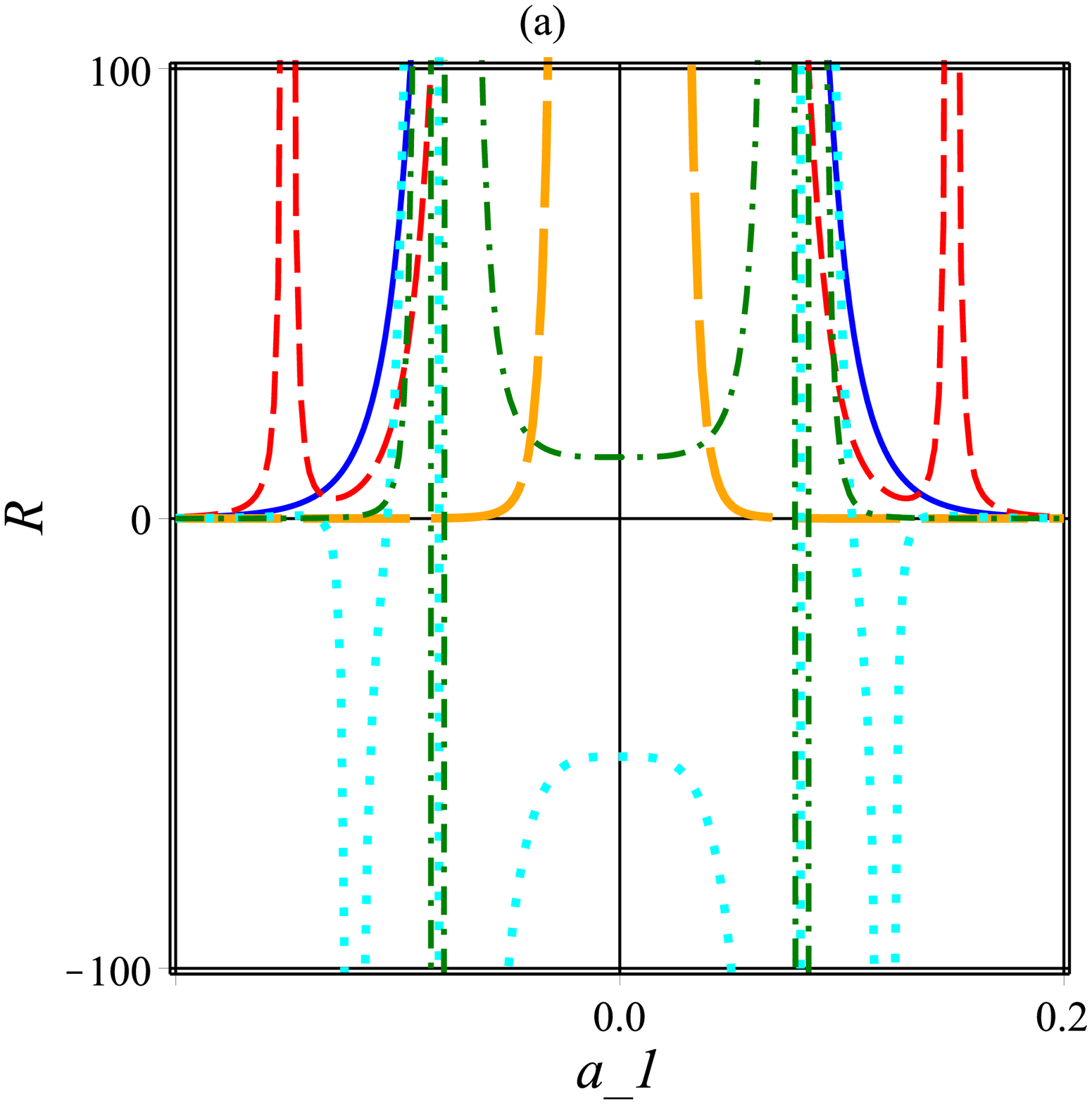}&\includegraphics[width=50 mm]{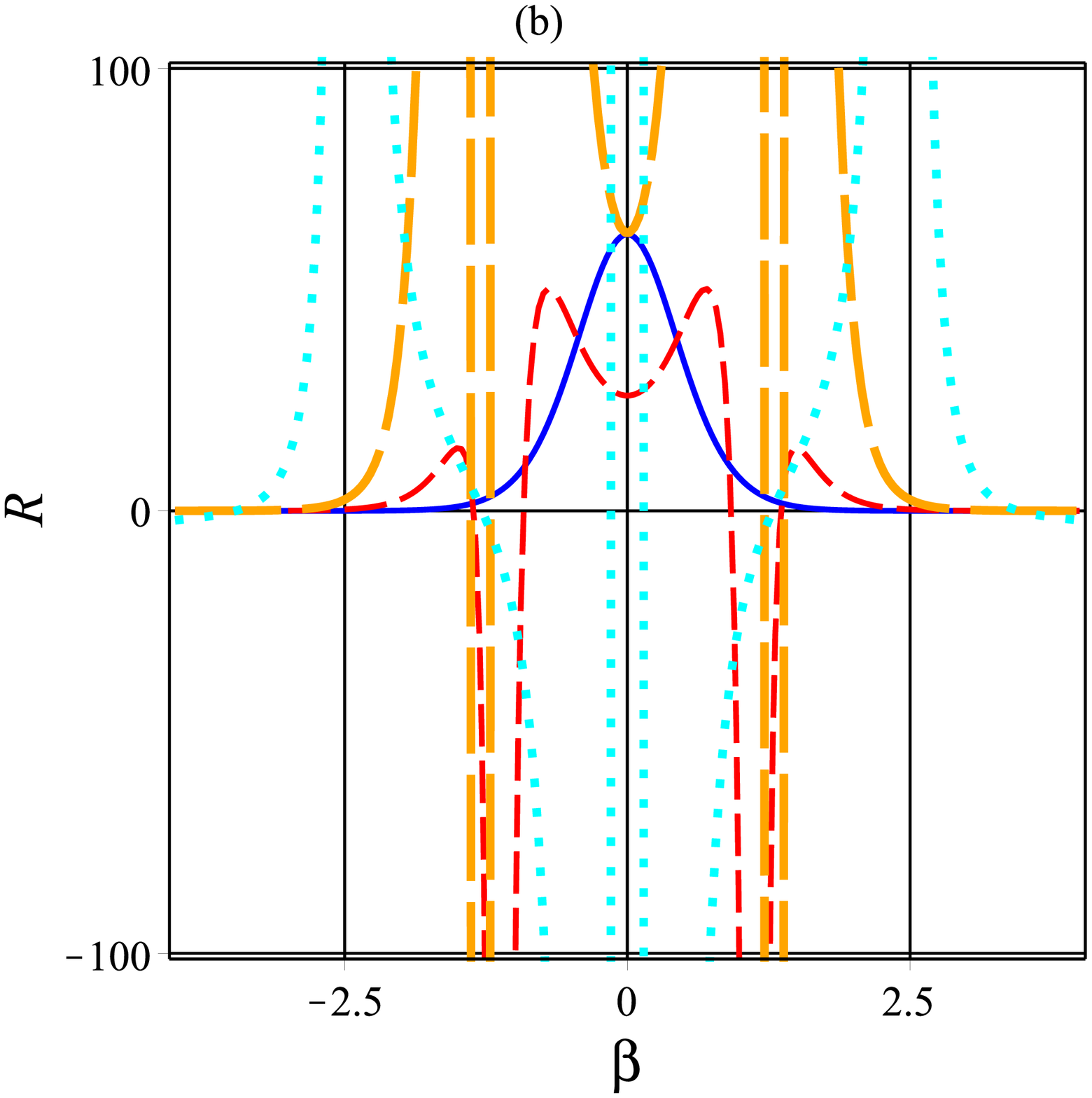}
 \end{array}$
 \end{center}
\caption{$R$ in terms of $a_{1}$ (a) and $\beta$ (b) for $\delta_{2}=-0.001$. (a) $\alpha=0$, $\beta=0$
(blue solid); $\alpha=1$, $\beta=0$ (red dash); $\alpha=0$, $\beta=2.4$ (orange long dash);
$\alpha=1$, $\beta=0.8$ (cyan dot); $\alpha=1$, $\beta=1.6$ (green dash dot). (b) $\alpha=0$, $a_{1}=-0.1$
(blue solid); $\alpha=1$, $a_{1}=-0.1$ (red dash); $\alpha=1$, $a_{1}=-0.09$ (orange long dash); $\alpha=1$, $a_{1}=-0.04$ (cyan dot).}
 \label{fig:5}
\end{figure}

It is possible to relate the microscopic degrees of freedom of a black object with a conformal field theory \cite{l1, 1503.07418}.
So, using this relation for a charged dilatonic black Saturn,   the
modular invariance of the partition
function of the conformal field theory  would constraint the entropy of the charged dilatonic black Saturn $S(\beta_{\kappa})$
\cite{card} to have the form $ S(\beta_{\kappa}) = a \beta_{\kappa}^l   + b \beta_{\kappa}^{-j}   $,
where  $l, j, a, b > 0$. The extremum of this function defines the equilibrium temperature as  $\beta_0 = (jb/la)^{1/ l+j} = T^{-1}$.
Expanding the entropy of the charged dilatonic black Saturn around this extremum, we obtain
\begin{eqnarray}
S(\beta_{\kappa}) &=& [(j/l)^{l/(l+j)} + (l/j)^{j/(l+j)} ](a^j b^l)^{1/(l+j)}
\nonumber \\
&&
+ \frac{1}{2}[(l+j) l^{(j+2)/(l+j)} l^{(l-2)/(l+j)}]
( a^{j+2}b^{l-2})^{{1}/(l+j)}\nonumber \\ && \times
(\beta_{\kappa} - \beta_0)^2, \label{a2}
\end{eqnarray}
Now, we can write
\begin{eqnarray}
S_0 &=& (j/l)^{l/(l+j)} + (l/j)^{j/(l+j)} (a^j b^l)^{{1}/(l+j)},
\nonumber \\
\left(\frac{\partial^2 S(\beta_{\kappa})}{\partial \beta_{\kappa}^2 }\right)_{\beta_{\kappa} = \beta_0} &=& (l+j) l^{(j+2)/(l+j)}
j^{(l-2)/(l+j)} ( a^{j+2}b^{l-2}  )^{1/(l+j)}.
 \end{eqnarray}
Thus, we obtain the values of $a, b$, and write
\begin{eqnarray}
\left(\frac{\partial^2 S(\beta_{\kappa})}{\partial \beta_{\kappa}^2 }\right)_{\beta_{\kappa} = \beta_0} =  \mathcal{Y} S_0 T^2~,
\end{eqnarray}
where
\begin{equation}
   \mathcal{Y} = \left[  \left(\frac{(l+j)
l^{(j+2)/(l+j)} j^{(l-2)/(l+j)}}{(j/l)^{l/(l+j)} + (l/j)^{j/(l+j)} }
 \right) \left(\frac{j}{l} \right)^{2/(l+j)}  \right].
\end{equation}
The factors $\mathcal{Y}$ which is independent of the
parameters in the charged dilatonic black Saturn   can be absorbed using some redefinition, just as it was done for other black objects
\cite{l1, 1503.07418}.
So, using  the relation between the corrections to the entropy and a conformal field theory, we obtain  \cite{l1, 1503.07418},
\begin{equation}\label{TF2}
S = S_{0} -\frac{\alpha}{2} \ln |S_{0}T^{2}|+\cdots,
\end{equation}
where the  variable $\alpha$ is again introduced to parameterize the effect of thermal fluctuations on the thermodynamics
of charged dilatonic black Saturn.
In this paper, we will study corrected thermodynamics using both relations given by (\ref{TF1}) and (\ref{TF2}).
We will demonstrate that both these expressions lead to the same physical results. Hence, the effect of corrections to the entropy obtained
from the fluctuation  of energy  are  the same as the effects of  corrections to the entropy obtained using a conformal field theory.
Now using   equation (\ref{TF2}),  we can write the internal energy as follow,
\begin{equation}\label{TF13}
E=\int{T dS}=E_{0}-\frac{\alpha}{2}\ln{S_{0}T^{2}},
\end{equation}
where $E_{0}$ given by the equation (\ref{TF5}). Also we can also write
\begin{equation}\label{TF14}
F=F_{0}-\frac{\alpha}{2}(1-T)\ln{S_{0}T^{2}},
\end{equation}
where $F_{0}$ given by the equation (\ref{TF10}). So, we can express the specific heat as
\begin{equation}\label{TF15}
C=C_{0}-\alpha(1+\frac{C_{0}}{2S_{0}}).
\end{equation}
 The specific heat obtained here is similar to the specific heat obtained before in   (\ref{TF12-1}). This also means that
 conformal field theory can be used to analyse the
   thermodynamical stability of this system. This is because of the similarity between the equations (\ref{TF15}) and (\ref{TF12-1}).
We can also  show that both (\ref{TF1}) and (\ref{TF2}) yield to the similar result. Now we can write corrected entropy as
\begin{equation}\label{TF16}
S={\frac {{a_{1}}^{4}\cosh{\beta} }{\delta_{2}}}-\frac{1}{2}\alpha \ln
 \left( {\frac {-2{\delta_{1}}^{2}}{\delta_{2}\cosh{\beta} }}
 \right),
\end{equation}
where the values of  $\delta_{1}$ and $\delta_{2}$ are fixed. Now we can varies the values of   $a_{1}$ and $\beta$. We can see that the
only difference with the equation (\ref{TF7}) is that $-\frac{\alpha}{2}\ln2$, and this does not generate any important physical
effect.  So, we have demonstrated that the corrections    obtained from the fluctuations in the energy are similar to the corrections
  obtained using a conformal field theory.\\
  It may be noted that
  it is possible to study the phase transition of this thermodynamical system  \cite{Geo1}.
In the geometro-thermodynamic formalism \cite{Geo2, Geo3},  the  thermodynamic metric given by,
\begin{equation}\label{Geo1}
g=\left(E^{a}\frac{\partial S}{\partial E^{a}}\right)\left(\eta_{b}^{c}\frac{\partial^{2}S}{\partial E^{c}\partial E^{d}}dE^{b}dE^{d}\right),
\end{equation}
where $\eta_{b}^{c}=(-1, 1, 1,\cdots, 1)$ at the equilibrium. Here $E^{a}$ are the relevant extensive parameters of the system.
For the charged dilatonic black Saturn, these are the black hole charges $\beta$ and $a_{1}$. Therefore the thermodynamic metric reduced to,
\begin{equation}\label{Geo2}
g=g_{11}da_{1}^{2}+g_{22}d\beta^{2}=\left(a_{1}\frac{\partial S}{\partial a_{1}}+\beta\frac{\partial S}{\partial \beta}\right)
\left(-\frac{\partial^{2}S}{\partial a_{1}^{2}}da_{1}^{2}+\frac{\partial^{2}S}{\partial \beta^{2}}d\beta^{2}\right).
\end{equation}
Using the equation (\ref{TF16}), we can obtain,
\begin{eqnarray}\label{Geo3}
\delta_{2}^{2}g_{11}&=&-6\left(2a_{1}^{4}\beta\cosh{\beta}\sinh{\beta}+8a_{1}^{4}\cosh^{2}{\beta}+\alpha\beta\delta_{2}\sinh{\beta}\right) a_{1}^{2},\nonumber\\
4\delta_{2}^{2}\cosh^{3}{\beta}g_{22}&=&4a_{1}^{8}\beta\cosh^{4}{\beta}\sinh{\beta}+16a_{1}^{8}\cosh^{5}{\beta}+2a_{1}^{4}\alpha\beta\delta_{2}\cosh^{3}{\beta}\sinh{\beta}\nonumber\\
&+&2a_{1}^{4}\alpha\beta\delta_{2}\cosh{\beta}\sinh{\beta}+8a_{1}^{4}\alpha\delta_{2}\cosh^{2}{\beta}+\beta\alpha^{2}\delta_{2}^{2}\sinh{\beta}.
\end{eqnarray}
So,  the second law of thermodynamics \cite{Geo4}, can be expressed as
\begin{equation}\label{Geo4}
\frac{\partial^{2}S}{\partial a_{1}^{2}}+\frac{\partial^{2}S}{\partial \beta^{2}}\geq0.
\end{equation}
Using the equations (\ref{TF16}) and (\ref{Geo4}), we obtain,
\begin{equation}\label{Geo5}
\frac{2a_{1}^{2}(12+a_{1}^{2})\cosh^{3}{\beta}+\alpha\delta_{2}}{2\delta_{2}\cosh^{2}}\geq0.
\end{equation}
The second law of thermodynamics is satisfied by taking negative and infinitesimal values of $a_{1}$ and $\delta_{2}$
along with an appropriate choice of $\beta$ (for example,  $\alpha=1$, $a_{1}\geq-0.00650$ and $\delta_{2}=-0.001$).
It is clear that with negative $\delta_{2}$, presence of logarithmic corrections are necessary to verify
the second law of thermodynamics. For instance, the following values verify the equation
(\ref{Geo5}), $\alpha=1$, $a_{1}=-0.001$, $-2\leq\beta\leq2$ and $\delta_{2}=-0.001$.\\
Now, thermodynamic interaction can be calculated from scalar curvature of the metric i.e., the  Ricci scalar $R$.
Analytic expression of $R$ for the metric (\ref{Geo2}) is complicated, so we will discuss   it graphically.
However,  we can write explicit expression for some special cases.\\
If we set $\alpha=\beta=0$ (uncorrected, uncharged) then find,
\begin{equation}\label{Geo6}
R=\frac{5\delta_{2}^{2}}{8a_{1}^{8}}.
\end{equation}
It is always positive quantity.\\
If we set $\beta=0$ (uncharged) then find,
\begin{equation}\label{Geo7}
R=\frac{(60a^{8}+36a^{4}\alpha\delta_{2}+7\alpha^{2}\delta_{2}^{2})\delta_{2}^{2}}{24(4a^{8}+4a^{4}\alpha\delta_{2}+\alpha^{2}\delta_{2}^{2})a_{1}^{8}},
\end{equation}
which may be negative or positive depend on the values of $\alpha$, $a_{1}$ and $\delta_{2}$.\\
If we set $\alpha=0$ (uncorrected) then find,
\begin{equation}\label{Geo8}
R=\frac{\delta_{2}^{2} \left(-\cosh^{4}{\beta}+8\beta\cosh{\beta} \sinh{\beta}+41\cosh^{2}{\beta}-\beta^{2} \right) }
{\cosh^{3}{\beta}a_{1}^{8} \left(  (12 \beta^{2}+64) \cosh^{3}{\beta}+\beta \sinh{\beta} ({\beta}^{2}+48)\cosh^{2}{\beta}
-12\beta^{2}\cosh{\beta}-\beta^{3}\sinh{\beta}\right) }.
\end{equation}
For the general case, we have plots of the Fig. \ref{fig:5}. We can observe the  phase transition for special values of parameters.
In the simplest case, when  $\alpha=\beta=0$,  we can see divergency of curvature around $a_{1}=0$ (blue solid line of the
Fig. \ref{fig:5} (a)). Furthermore, when  $\alpha=0$ (uncorrected),  we can see $R\geq0$ everywhere.

\section{Conclusions}
In this paper,  we have analysed the corrections to the thermodynamics of a
charged dilatonic black Saturn.
Using numerical analysis,  we found simple expression for the temperature and entropy of this charged dilatonic black Saturn.
  We analysed the effects of thermal fluctuations on the thermodynamics of charged black Saturn
using two different methods. Thus, we first analysed the effect of thermal fluctuations by relating the fluctuations in the entropy of
this charged dilatonic black Saturn to
the fluctations in its energy.  Then we analysed the fluctations in the entropy using the relation
between the entropy of this charged dilatonic black Saturn and a conformal field theory. It was demonstrated that similar physical results
were obtained from both these methods. We also analysed the effect such thermal fluctuations will have on the thermodynamic stability
of this charged dilatonic black Saturn.  The validity of the second law of thermodynamics was investigated using this formalism.\\
It may be noted thermal instability and thermodynamic geometry of topological dilaton black holes coupled to nonlinear
electrodynamics has been analyzed \cite{p1}. In this analysis  the stability analysis was performed in both canonical
and grand canonical ensembles. The
phase transition and thermodynamic geometry of Einstein-Maxwell-dilaton black holes has also been discussed \cite{p2}. It was observed that this system
can have  three different critical behaviors near the critical points for these black holes.  A thermodynamical metric was used
for analyzing the  thermodynamical geometry of these black holes. Magnetically charged regular black hole in a model of nonlinear
electrodynamics have also been studied \cite{p4}. The heat capacity at constant charge was used for analyzing the stability of these
black holes. The thermodynamics of rotating thin shells has been studied in  the BTZ space-time \cite{p5}
The topological black hole solutions have been studied using a third order Lovelock Ads black holes in the presence of nonlinear electrodynamics \cite{p6}.
It was demonstrated that the thermodynamic quantities of the black hole solutions satisfy the first law of thermodynamics.
It would be interesting to analyze the corrections to these black holes from thermal fluctuations. We expect that the corrections to the entropy will
have interesting consequences for the physics of all these black holes.

\end{document}